# Pressure-induced superconductivity reentrant in transition metal dichalcogenide TiSe$_2$

Wei Xia,[1, 3, *] Jiaxuan Wu,[2, *] Zhongyang Li,[1, 4] Jian Yuan,[1] Chao An,[5] Xia Wang,[7] Na Yu,[7] Zhiqiang Zou,[7] Gang Liu,[4] Chunyin Zhou,[8] Jiajia Feng,[4] Lili Zhang,[8] Zhaohui Dong,[8] Bin Chen,[4] Zhaorong Yang,[6] Zhenhai Yu,[1, †] Hanghui Chen,[2, 9, ‡] and Yanfeng Guo[1, 3, §]

[1] School of Physical Science and Technology, ShanghaiTech University, Shanghai 201210, China
[2] NYU-ECNU Institute of Physics, NYU Shanghai, Shanghai 200122, China
[3] ShanghaiTech Laboratory for Topological Physics, ShanghaiTech University, Shanghai 201210, China
[4] Center for High Pressure Science and Technology Advanced Research, Shanghai, 201203
[5] Information Materials and Intelligent Sensing Laboratory of Anhui Province, Institutes of Physical Science and Information Technology, Anhui University, Hefei 230601, China
[6] Anhui Province Key Laboratory of Condensed Matter Physics at Extreme Conditions, High Magnetic Field Laboratory, Chinese Academy of Sciences, Hefei, 230031 China
[7] Analytical Instrumentation Center, School of Physical Science and Technology
[8] Shanghai Synchrotron Radiation Facility, Shanghai Advanced Research Institute, Chinese Academy of Sciences, Shanghai 201204, China
[9] Department of Physics, New York University, New York, NY 10012, USA

**Through either elements intercalation or application of pressure, transition metal dichalcogenide 1$T$-TiSe$_2$ exhibits superconductivity in proximity to a charge density wave (CDW) quantum critical point (QCP), thus providing an ideal avenue to study the correlation between the two symmetry-breaking exotic quantum electronic states. We report herein that, in addition to the well-known superconducting dome that emerges within the low pressure range of 2 - 4 GPa and peaks with the maximal $T_c$ of about 1.8 K, the pressure induces another separate superconducting transition** starting around 15 GPa with a substantially higher $T_c$ that reaches 5.6 K at about 21.5 GPa. The high-pressure X-ray diffraction and Raman spectroscopy measurements unveil that the superconductivity reentrant is caused by a first-order structural phase transition (from $P\bar{3}m1$ space group to *Pnma* space group), which is also supported by the density functional theory calculation. A comparative theoretical calculation also reveals that the conventional phonon-mediated mechanism can account for the superconductivity of 1*T*-TiSe$_2$ under low pressure, while the electron-phonon coupling of 4*O*-TiSe$_2$ under high pressure is too weak to induce the superconductivity with a $T_c$ as high as 5.6 K. This implies that the emergent superconductivity in the 4*O*-TiSe$_2$ may have an unconventional origin. Our finding would open a new window toward the discovery of more exotic quantum states in transition metal dichalcogenides via high pressure.

*Equal contributions

Corresponding authors:

[†] yuzhh@shanghaitech.edu.cn (ZHY),

[‡] hanghui.chen@nyu.edu (HHC),

[§] guoyf@shanghaitech.edu.cn (YFG)

The entanglement of charge density wave (CDW) state with many other intriguing phenomena such as superconductivity and magnetism, etc. [1-4], renders CDW materials of particular interest for exploring exotic physics properties. The CDW order in cuprate superconductors, for example, has been convinced to play a crucial role for elucidating the high superconducting critical temperature ($T_c$) superconductivity [1, 5-13]. In recent decades, the attentions on the correlation between superconductivity and CDW have also always been arrested by many CDW materials such as the transition-metal dichalcogenides (TMDs). With the simple chemical formula MX$_2$ where M = Nb, Ti, Ta, Mo, and X = S, Se, and Te, TMDs serve as a model system to study the interplay between CDW and superconductivity [14]. For example, trigonal 1$T$-TiSe$_2$ (space group: $P\bar{3}m1$, No.164) is a prototype of TMDs, which exhibits a CDW order below $T_{\text{CDW}}$ of 202 K under ambient pressure [15, 16] and superconductivity either by intercalation of foreign atoms [17, 18] or by applying pressure [19, 20]. Interestingly, the simultaneous application of Cu intercalation and pressure can shift the CDW quantum critical point (QCP). The highest $T_c$ is found to be pinned to the QCP, hinting at a strong correlation between the two electronic states. Moreover, manipulating the CDW state can also produce other exotic phenomena. For example, via inverting the periodic lattice distortion of the three-dimensional CDW state sectionally by using femtosecond laser, macroscopic domain walls of a transient two-dimensional ordered electronic state emerge and is associated with remarkably enhanced density of states near the surface, thus providing the possibility to explore unusual low-dimensional superconductivity [21]. These results strongly motivate the use of versatile techniques to create exotic states in 1$T$-TiSe$_2$.

Pressure can create superconductivity in a material either by inducing a structure phase transition or by modifying its electronic band structure. There have been numerous reports on pressure-induced superconductivity in various quantum materials such as the topological nodal-line semimetal SrAs$_3$ [22], iron-based and Cr-based superconductors [23, 24]. 1$T$-TiSe$_2$ has been studied under high pressure up to 10 GPa, showing that the pressure can suppress the CDW phase and induce superconductivity (denoted as SC-I phase). The $T_c$ exhibits a "dome-like" feature in the range of 2 - 4 GPa [19], with the maximum value reaching about 1.8 K. The high-pressure X-ray scattering study unveils a CDW QCP behavior at around 5.1 GPa [2]. Sincere there have only handful of work [25], a systematic study of 1$T$-TiSe$_2$ under hydrostatic pressure higher than 10 GPa is expected to provide opportunities for discovery of interesting properties.

In this work, by using the electrical transport measurements on 1$T$-TiSe$_2$ under pressure up to 21.5 GPa, we discover a superconductivity reentrant (denoted as SC-II phase) driven by a structure phase transition. The $T_c$ of SC-II phase reaches as high as 5.6 K at about 21.5 GPa and shows no any sign of saturation or decline. The high-pressure X-ray diffraction (XRD) and Raman spectroscopy measurements up to 30 GPa reveal that TiSe$_2$ undergoes a first-order structure transition from 1$T$ (space group $P\bar{3}m1$) to 4$O$ (space group $Pnma$) phase. First-principles calculations support this pressure-driven structural phase transition. A comparative *ab initio* calculation of electron-phonon properties of TiSe$_2$ finds that under low pressure, the conventional phonon-mediated mechanism can account for the experimentally observed superconductivity in the $1T$ phase; however, under high pressure, the electron-phonon coupling of $4O$-TiSe$_2$ is too weak to induce superconductivity with a $T_c$ as high as 5.6 K, implying that the origin of SC-II phase might be unconventional.

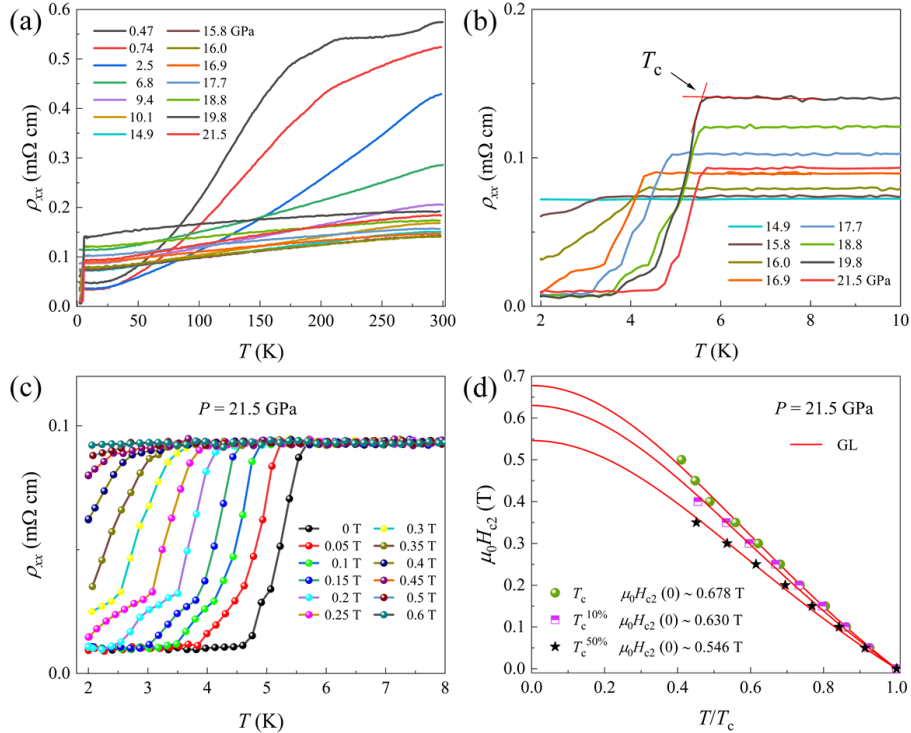

**Fig. 1.** Temperature dependent (a)-(b) resistivity for $1T$-TiSe$_2$ under various pressures. (c) Temperature dependence of resistivity under different magnetic fields at 21.5 GPa, showing that the increasing magnetic field gradually suppresses the superconducting transition. (d) Temperature dependence of the upper critical field $\mu_0 H_{c2}$ at 21.5 GPa. The critical magnetic fields correspond to the $T_c$, and 10% ($T_c^{10\%}$) and 50% ($T_c^{50\%}$) drop of the superconducting transition, respectively. The red lines are the fitting results by using the Ginzburg–Landau equation expressed as $\mu_0 H_{c2}(T) = \mu_0 H_{c2}(0)(1 - (T/T_c)^2)/(1 + (T/T_c)^2)$.

Fig. 1(a) depicts the temperature dependence of the longitudinal resistivity $\rho_{xx}$ of $1T$-TiSe$_2$ under various pressures up to 21.5 GPa. At ambient pressure, the influence of CDW order on $\rho_{xx}$ is clearly visible around 200 K, which eventually disappears when the pressure is up to 2.5 GPa. Limited by the measured low temperature, the superconductivity between 2 - 4 GPa with the maximum $T_c$ of 1.8 K was not observed [19]. In our experiment, no superconducting transition was observed till 14.9 GPa, as shown in Fig. 1(b). When the pressure $P$ is larger than 15.8 GPa, $\rho_{xx}$ displays a significant decrease at low temperature. At 15.8 GPa, a superconducting transition with $T_c$ of 3.23 K appears, and the $T_c$ increases to 5.6 K at 21.5 GPa. It is noted that the $T_c$ is determined from the crossing point of two straight lines above and below the superconducting transition. The 5.6 K $T_c$ of SC-II phase is much higher than that of SC-I phase (1.8 K), which resembles the cases in $(Li_{1-x}Fe_x)OHFe_{1-y}Se$ and K-Fe-Se [26, 27]. The superconducting transition is somewhat wide, due to the coexistence of both 1$T$-TiSe$_2$ and 4$O$-TiSe$_2$ phases in our measured pressure range, which will be discussed later. The magnetic field dependence of $\rho_{xx}$ is presented in Fig. 1(c), which displays gradually suppressed $T_c$ by the magnetic field. The behavior is usually viewed as a characteristic for the superconducting transition. The temperature dependence of the upper critical field ($\mu_0H_{c2}$) obtained at 21.5 GPa is plotted in Fig. 1(d) by using the Ginzburg–Landau(GL) formula $\mu_0H_{c2}(T) = \mu_0H_{c2}(0)(1 - (T/T_c)^2)/(1 + (T/Tc)^2)$. The red solid lines denoting the fitting results show the $\mu_0H_{c2}(0)$ values of about 0.678 T, 0.630 T, 0.546 T for $T_c$, $T_c^{10\%}$, and $T_c^{50\%}$, respectively. To obtain the Hall coefficient $R_H = d\rho_{xy}/dB$, the transverse resistivity $\rho_{xy}$ was measured and is presented in panels (a)-(b) of Fig. S3. The $R_H$ values were obtained as the slope of the linear fitting by employing the single band model to $\rho_{xy}$. $R_H$ displays a non-monotonic variation with pressure, which is plotted against the pressure in Fig. S3(c). The electron-type carrier density was thus estimated as $n_e = -1/(R_H e)$. As shown in Fig. S3(c), $n_e$ slightly decreases first and then takes almost a constant value of $\sim 3 \times 10^{28}$ m$^{-3}$ for $P <$ 15 GPa. Above 15 GPa, it starts to increase linearly up to $\sim 5 \times 10^{28}$ m$^{-3}$ at 18.8 GPa, nicely consistent with the evolution trend of $T_c$ ($P$) with the pressure. The results indicate that emergence of the SC-II phase with higher $T_c$ is associated with a concurrent enhancement of electron carrier density.

Based on the electrical transport measurements under pressure, the temperature-pressure phase diagram of 1$T$-TiSe$_2$ is established, as shown in Fig. 2(a). It should be noted that the CDW temperature and superconducting dome (×10) of SC-I phase are taken from Ref. [19]. The maximum $T_c$ (×10) of SC-II phase is the highest among the values of both intercalated and pressurized TiSe$_2$ since the previous record is 4.15 K in Cu$_{0.08}$TiSe$_2$ [18]. Such case was also observed in pressurized (Li$_{1-x}$Fe$_x$)OHFe$_{1-y}$Se and K-Fe-Se superconductors [26, 27]. However, in most superconductors such as KMo$_3$As$_3$, the $T_c$ of the pressure-induced superconductivity reentrant usually has a lower $T_c$ than that of the initial SC-I phase [28]. A prominent feature of the re-emergent superconductivity is that the $T_c$ is almost robust against our measured pressure, indicating a robust superconductivity or a much wider superconducting dome.

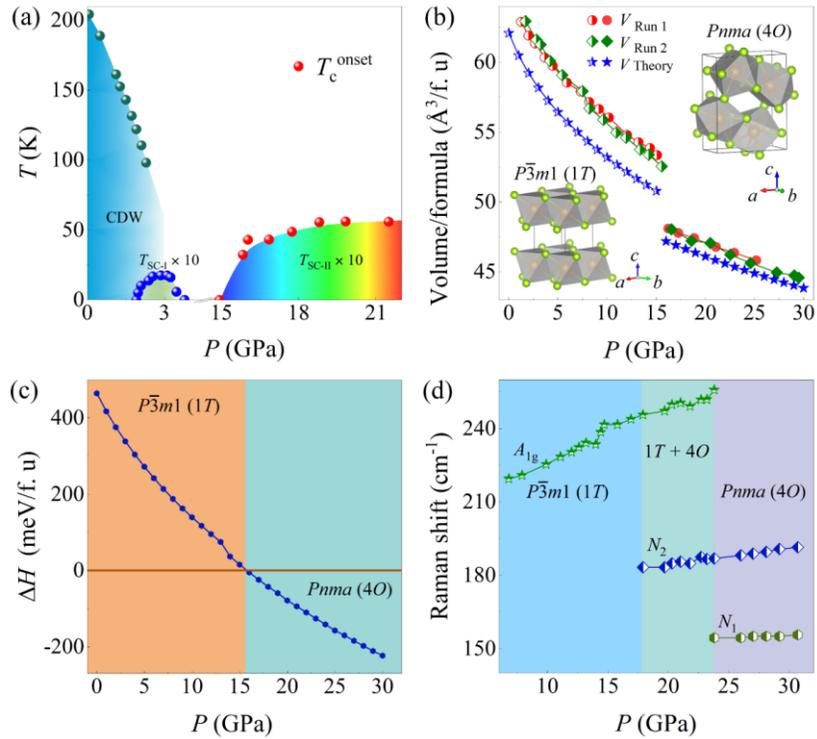

**Fig. 2.** (a) Pressure-temperature phase diagram of 1$T$-TiSe$_2$, which shows the evolution of the CDW and $T_c$ (×10) against pressure. The CDW temperature against pressure and the superconducting dome of SC-I (×10) are taken from Ref. [19]. (b) Experimental and calculated pressure dependence of relative volumes for 1$T$ and 4$O$ phases. The schematic crystal structures for both phases are shown for a comparison. (c) The DFT-calculated total enthalpy difference Δ$H$ between 4$O$ and 1$T$ phases as a function of hydrostatic pressure. A first-order structural phase transition from 1$T$ phase to 4$O$ phase occurs around 16 GPa, which is in good agreement with the experiment. (d) Pressure dependence of Raman shift of phonon modes ($A_{1g}$, $N_1$ and $N_2$).

The TMDs can adopt diverse crystal structures, namely, the 1$T$, 2$H$, and 3$R$ phases, etc., where the Arabic numerals denote the quantity of formulas in one unit cell and $T$, $H$, and $R$ are the abbreviations for the trigonal, hexagonal and rhombohedral structures, respectively. At ambient conditions, TiSe$_2$ crystallizes into the trigonal structure with space group $P\bar{3}m1$, which is namely the 1$T$-TiSe$_2$ phase where Ti and se atoms locate at the 1a (0, 0, 0) and 2d (0.333, 0.667, 0.255) Wyckoff positions, respectively [29]. The phase has a layered structure with edge-connected TiSe$_6$ octahedron layers running along the $c$-axis with weak van der Waasl interaction between adjacent layers. The AD-XRD patterns (Run 1 experiment) of 1$T$-TiSe$_2$ under selected pressures are presented in Fig. 3(a), showing no appreciable variation below ∼15.0 GPa except a gradual shift of the Bragg peaks toward higher angles. Result of another independent AD-XRD (experiment Run 2) measurements on 1$T$-TiSe$_2$ is presented in Fig. S2(c), which is nicely consistent with that of experiment Run 2. The XRD patterns below ∼ 15.0 GPa were identified with the $P\bar{3}m1$ phase using Rietveld refinement, seen in Fig. S2(b). With further increasing the pressure above 15.0 GPa, new diffraction peaks appear, hinting the formation of a new structure. The intensity of the new diffraction peaks is apparently enhanced as the pressure increases. However, the structure analysis revealed that the $P\bar{3}m1$ phase and the high-pressure phase can coexist within our measured pressure range. The $P\bar{3}m1$ phase was recovered after released the sample to ambient pressure, indicating that the pressure-induced structural phase transition was reversible, as shown in Fig. S2(a).

The structure analysis of the high-pressure phase of TiSe$_2$ was referred to several previous work [30–35], which allowed us the choice of the subgroups of $P\bar{3}m1$, $Pnma$ and $P\bar{6}2m$ as the initial models for refinements. The cotunnite-type $Pnma$ structure was found to yield the best refinement results. Since there are 4 formulas in one unit cell for the $Pnma$ phase, the high-pressure phase of TiSe$_2$ was eventually assigned as the 4$O$ phase according to the structure designation in TMDs. The coordination numbers (CN) for Ti atoms in 1$T$ and 4$O$ phases are 6 and 8, respectively, revealing that the pressure can increase the CN. In addition, TiSe$_6$ octahedra are connected by sharing edges in 1$T$-TiSe$_2$, while the TiSe$_8$ hendecahedra share faces in the 4$O$ phase. The relative volumes of the 1$T$ and 4$O$ phases as a function of pressure are summarized in Fig. 2(b). It should be noted that the pressure-induced structural phase transition is accompanied with the occurrence of superconductivity reentrant. To further verify our structure analysis, the Raman scattering spectroscopy was employed to characterize the crystal structure of TiSe$_2$ under pressure. There are nine zone center vibrational modes for $1T$-TiSe$_2$ according to group analysis. $\Gamma = A_{1g} + E_g(2) + 2A_{2u} + 2E_u(2)$. Here, the gerade ($E_g$ and $A_{1g}$) modes are Raman active and the ungerade ($E_u$ and $A_{2u}$) modes are infrared-active. The Raman active $A_{1g}$ mode represents the stretching of two Se atoms moving relatively to one another parallel to $c$-axis and $E_g$ mode represents the symmetric in-plane bending of the Se atoms [36, 37]. Panels (b)-(c) of Fig. 3 show the Raman spectra at selected pressures, which display a monotonous increase (hardening) in Raman shift as the pressure increases, consistent with the Raman scattering spectroscopy measurements results in a previous study [38]. It is visible that new Raman peaks, marked with $N_1$ and $N_2$, appear above 17.9 GPa, demonstrating the pressure-induced structural phase transition again and being consistent with the high pressure AD-XRD results [25, 38]. The most intense Raman peak of $1T$-TiSe$_2$, namely $A_{1g}$, is observed both in $1T$ and the high-pressure phase, signifying the coexistence of both phases at high pressure, which is also observed in the AD-XRD measurements. Furthermore, the $A_{1g}$ peak gradually decreases in intensity and eventually disappears above ∼ 24 GPa. Fig. 2(d) presents the Raman shift of 1$T$-TiSe$_2$ versus pressure, showing a monotonous increase of Raman modes $A_{1g}$, $N_1$ and $N_2$ in Raman shift with increasing pressure. The pressure-induced phase transition from $P\bar{3}m1$ phase to *Pnma* phase is marked by different background colors in the figure.

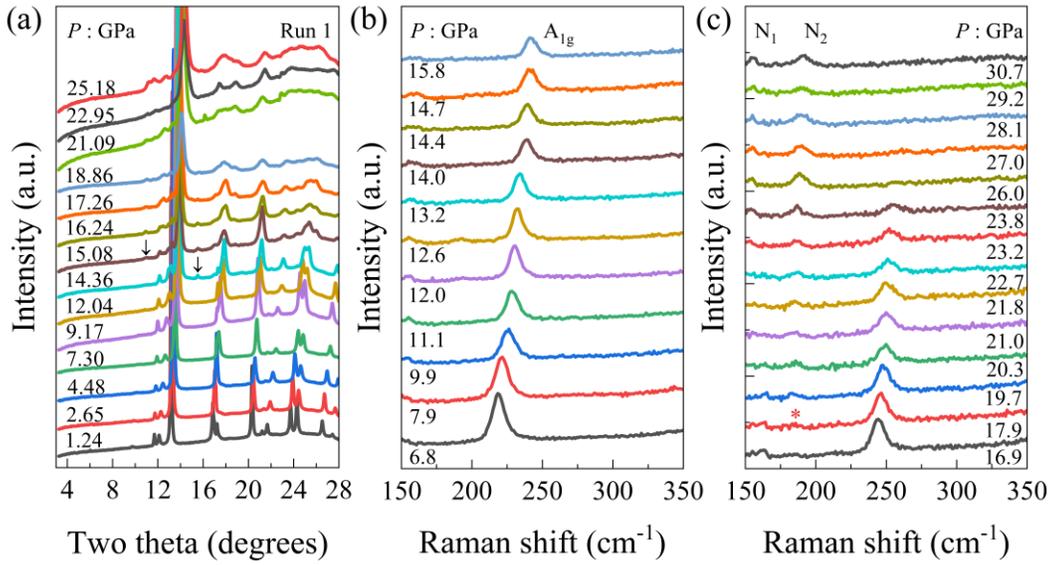

**Fig. 3.** (a) The selected room temperature AD-XRD patterns of $1T$-TiSe$_2$ under various pressures up to 25.2 GPa (Run 1). (b) and (c) Pressure dependence of Raman spectra of $1T$-TiSe$_2$ using 4:1 methanol ethanol mixture.

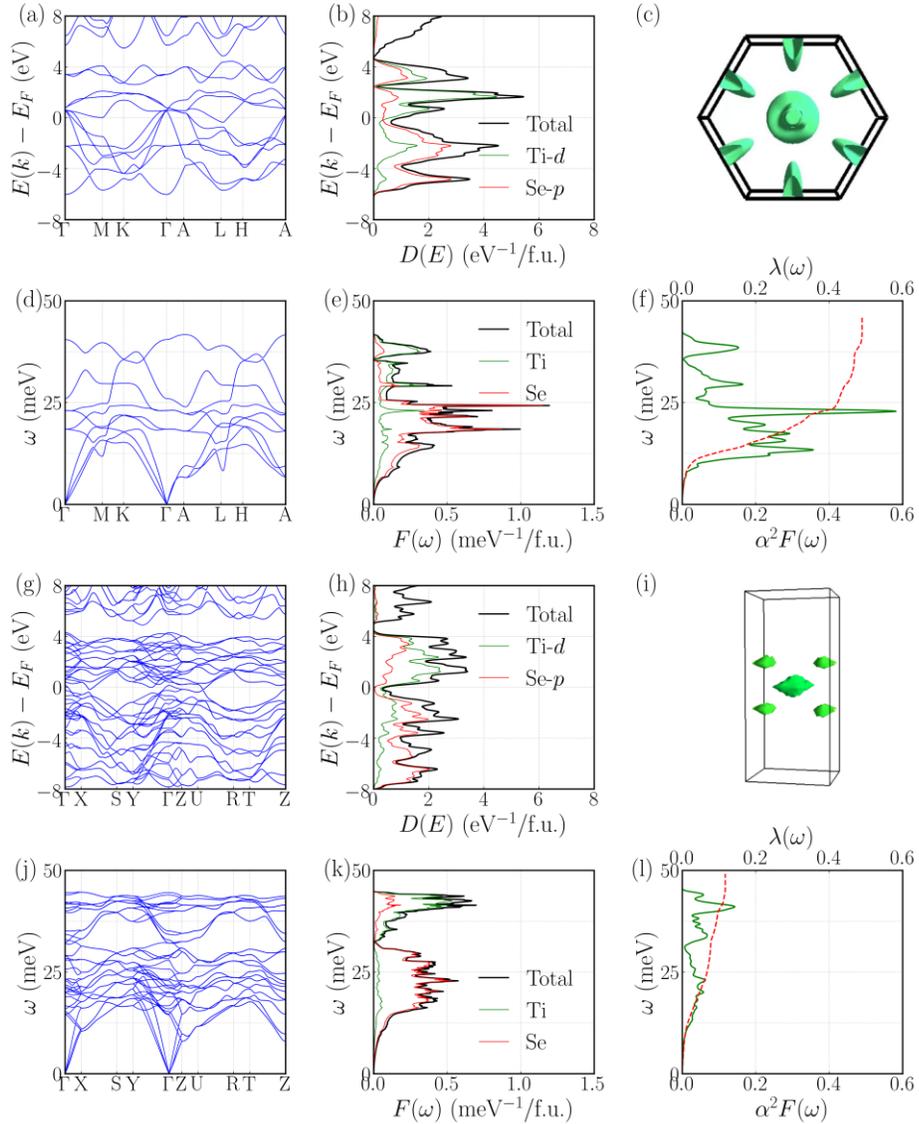

**Fig. 4.** (a)-(f): Electron and phonon properties of 1$T$-TiSe$_2$. (a) Electronic band structure of 1$T$-TiSe$_2$. (b) Electron density of states of 1$T$-TiSe$_2$. The black, green and red curves are total, Ti-d projected and Se-p projected densities of states, respectively. (c) Fermi surface of 1$T$-TiSe$_2$. (d) Phonon band

structure of 1$T$-TiSe$_2$. (e) Phonon density of states of 1$T$-TiSe$_2$. The black, green and red curves are total, Ti projected and Se projected densities of states, respectively. (f) Electron-phonon spectral function $\alpha^2 f(\omega)$ (green curve) and accumulative electron-phonon coupling $\lambda(\omega)$ (red curve) of 1$T$-TiSe$_2$. (g)-(l): electron and phonon properties of 4$O$-TiSe$_2$. (g) Electron band structure of 4$O$-TiSe$_2$. (h) Electron density of states of 4$O$-TiSe$_2$. The black, green and red curves are total, Ti-d projected and Se-p projected densities of states, respectively. (i) Fermi surface of $Pnma$-TiSe$_2$. (j) Phonon band structure of $Pnma$-TiSe$_2$. (k) Phonon density of states of 4$O$-TiSe$_2$. The black, green and red curves are total, Ti projected and Se projected densities of states, respectively. (l) Electron-phonon spectral function $\alpha^2 f(\omega)$ (green curve) and accumulative electron-phonon coupling $\lambda(\omega)$ (red curve) of 4$O$-TiSe$_2$

To gain in-depth insights into the nature of superconductivity in TiSe$_2$, we performed a comparative *ab initio* study on the 1$T$ (space group $P\bar{3}m1$) and 4$O$ (space group Pnma) phases. We first studied the 1$T$-TiSe$_2$ under 3 GPa where CDW is completely suppressed. Panels (a)-(c) of Fig. 4 show the electronic band structure, electron density of states and Fermi surface of 1$T$-TiSe$_2$ under 3 GPa, respectively. We find that close to the Fermi level, there is a strong hybridization between Ti-$d$ states and Se-$p$ states. Panels (d)-(f) of Fig. 4 show phonon band structure, phonon density of states, electron-phonon spectral function $\alpha^2 F(\omega)$ and accumulative electron-phonon coupling $\lambda(\omega)$ of 1$T$-TiSe$_2$ under 3 GPa, respectively. Under a pressure of 3 GPa, the $P\bar{3}m1$ structure is free of imaginary phonon modes, indicating that the CDW is completely suppressed. The total electron-phonon coupling $\lambda$ is about 0.5, which is sufficient to induce phonon-mediated superconductivity of about 2 K as estimated by McMillian equation using the Morel-Anderson pseudopotential $\mu^* = 0.1$. The theoretical result is in good agreement with the experiment (see Fig. 2). Next, using the same method, we studied 4$O$-TiSe$_2$ under 20 GPa in which superconductivity re-emerges in experiment. Panels (g)-(i) of Fig. 4 show electronic band structure, electron density of states and Fermi surface of 4$O$-TiSe$_2$ under 20 GPa. The hybridization between Ti-$d$ states and Se-$p$ states is still substantial in the $Pnma$ structure, but the Fermi surface of 4$O$-TiSe$_2$ is very different from that of 1$T$-TiSe$_2$ under 3 GPa because of dissimilar crystal structures. Panels (j)-(l) of Fig. 4 show phonon band structure, phonon density of states, electron-phonon spectral function $\alpha^2 F(\omega)$ and accumulative electron-phonon coupling $\lambda(\omega)$ of 4$O$-TiSe$_2$ under 20 GPa, respectively. The *Pnma* structure is dynamically stable, i.e. exhibits no imaginary phonon modes. However, the total electron-phonon coupling $\lambda$ of 4*O*- TiSe$_2$ under 20 GPa is only 0.2, smaller than the $\lambda$ of 1*T*-TiSe$_2$ under 3 GPa. This indicates that the conventional electron-phonon mechanism alone is not sufficient to account for the emergent superconductivity in 4*O*-TiSe$_2$ under high pressure, which has a $T_c$ as high as 5.6 K. The discrepancy between the electron-phonon coupling and the high $T_c$ of SC-II suggests the possibility of unconventional superconductivity in 4*O*-TiSe$_2$, which deserves further study.

In summary, we observe a pressure-driven superconductivity reentrant in TiSe$_2$ above 15.8 GPa, with $T_c$ reaching as high as 5.6 K and no sign of saturation or decline up to 21.5 GPa. High-pressure synchrotron x-ray diffraction and Raman spectroscopy measurements identify a new 4*O* crystal structure of TiSe$_2$ (space group *Pnma*) under high pressure, which leads to the re-emerging of superconductivity. The 1*T* to 4*O* structural phase transition is not complete within the measured pressure range, so the two structures may coexist under our measured pressure. The $T_c$ of 4*O*-TiSe$_2$ is substantially higher than the previously reported $T_c$ of intercalated and pressurized 1*T*-TiSe$_2$. First-principles calculations support the pressure-driven first-order structure phase transition in TiSe$_2$ (from 1*T* to 4*O*). The calculated transition pressure is in good agreement with the experiment. A comparative study of electron-phonon properties of TiSe$_2$ finds that while conventional phonon-mediated mechanism may account for the superconductivity found in 1*T*-TiSe$_2$ under low pressure, the electron-phonon coupling of 4*O*-TiSe$_2$ is not large enough to produce the experimentally observed high superconducting transition temperature under high pressure. The discovery of super-conductivity reentrant in TiSe$_2$ points to a new direction of studying transition metal dichalcogenides, in particular clarifying the origin of SC-II will be in our future study.

Acknowledgments
The authors acknowledge the support by the National Natural Science Foundation of China (Grant Nos. 92065201, 11874264) and the Shanghai Science and Technology Innovation Action Plan (Grant No. 21JC1402000). Y. F. Guo acknowledges the start-up grant of ShanghaiTech University, the Program for Professor of Special Appointment (Shanghai Eastern Scholar) and the Open Project of Key Laboratory of Artificial Structures and Quantum Control (Ministry of Education), Shanghai Jiao Tong University (Grant No. 2020-04). H.C. is supported by the National Natural Science Foundation of China under project number 11774236. The authors also thank the support from the Analytical Instrumentation Center (Grant No. SPST-AIC10112914), SPST, ShanghaiTech University. We would like to thank Prof. Kai Liu for fruitful discussions.